\documentclass[runningheads]{llncs}
\usepackage{comment}
\usepackage{graphicx}
\graphicspath{ {./images/} }
\usepackage{multirow}
\usepackage{cite}
\usepackage{longtable}
\usepackage{xurl}
\begin{document}
\title{MAG-Net: Multi-task attention guided network for brain tumor segmentation and classification 
\thanks{All authors have contributed equally.}}
\titlerunning{MAG-Net}
%
\author{Sachin Gupta\and
Narinder Singh Punn\and
Sanjay Kumar Sonbhadra\and
Sonali Agarwal}

\authorrunning{Sachin et al.}

\institute{Indian Institute of Information Technology Allahabad, India\\
\email{\{mit2019075,pse2017002,rsi2017502,sonali\}@iiita.ac.in}}

\maketitle              
\begin{abstract}
Brain tumor is the most common and deadliest disease that can be found in all age groups. Generally, MRI modality is adopted for identifying and diagnosing tumors by the radiologists. The correct identification of tumor regions and its type can aid to diagnose tumors with the followup treatment plans. However, for any radiologist analysing such scans is a complex and time-consuming task. Motivated by the deep learning based computer-aided-diagnosis systems, this paper proposes multi-task attention guided encoder-decoder network (MAG-Net) to classify and segment the brain tumor regions using MRI images. The MAG-Net is trained and evaluated on the Figshare dataset that includes coronal, axial, and sagittal views with 3 types of tumors meningioma, glioma, and pituitary tumor. With exhaustive experimental trials the model achieved promising results as compared to existing state-of-the-art models, while having least number of training parameters among other state-of-the-art models.
\keywords{Attention \and Brain tumor \and Deep learning \and Segmentation.}
\end{abstract}
\section{Introduction}
Brain tumor is considered as the deadliest and most common form of cancer in both children and adults. Determining the correct type of brain tumor in its early stage is the key aspect for further diagnosis and treatment process. However, for any radiologist, identification and segmentation of brain tumor via multi-sequence MRI scans for diagnosis, monitoring, and treatment, are complex and time-consuming tasks.

Brain tumor segmentation is a challenging task because of its varied behavior both in terms of structure and function. Furthermore, the tumor intensity of a person differs significantly from each other. MRI is preferred over other imaging modalities~\cite{MRI} for the diagnosis of brain tumor because of its non-invasive property that follows from without the exposure to ionizing radiations and superior image contrast in soft tissues.

Deep learning has shown advancement in various fields with promising performance especially in the area of biomedical image analysis~\cite{punn2021modality}. The convolutional neural networks (CNN)~\cite{albawi2017understanding} are the most widely used models in image processing. The CNNs involve combination of convolution, pooling and activation layers accompanied with the normalization and regularization operations to extract and learn the target specific features for desired task (classification, localization, segmentation, etc.). In recent years various techniques have been proposed for identification (classification and segmentation) of the brain tumor using MRI images that achieved promising results~\cite{icsin2016review, zhou2019review}. However, most of the approaches use millions of trainable parameters that result in slower training and analysis time, while also having high variance in results in case of limited data samples.

In order to overcome the aforementioned drawbacks, Ronneberger et al.~\cite{ronneberger2015u} proposed U shaped network (U-Net) for biomedical image segmentation. The model follows encoder-decoder design with feature extraction (contraction path) and reconstruction phases (expansion path) respectively. In addition, skip connections are introduced to propagate the extracted feature maps to the corresponding reconstruction phase to aid upsample the feature maps. Finally, model produces segmentation mask in same dimensions as the input highlighting the target structure (tumor in our case). Following the state-of-the-art potential of the U-Net model, many U-Net variants are proposed to further improve the segmentation performance. Attention based U-Net model~\cite{oktay2018attention} is one such variant that tend to draw the focus of the model towards target features to achieve better segmentation results. The attention filters are introduced in the skip connections where each feature is assigned weight coefficient to highlight its importance towards the target features. Despite achieving the promising results, these models have millions of trainable parameter which can be reduced by optimizing the convolution operation. This can be achieved by incorporating the depthwise convolution operations~\cite{chollet2017xception} that is performed in two stages: depthwise and pointwise convlutions. The reduction in the number of the parameters and multiplications as compared to standard convolution operation can represented as $1/r + 1/f^2$, where $r$ is the depth of the output feature map and $f$ is the kernel height or width~\cite{punn2020chs}. The achieved reduction in number of parameters and multiplications is $\sim80\%$. Following this context, attention guided network is proposed that uses depthwise separable convolution for real time segmentation and classification of the brain tumor using MRI imaging. The major contribution of the present research work is as follows:
\begin{itemize}
    \item A novel model, Multi-task (segmentation and classification) attention guided network (MAG-Net) is proposed for brain tumor diagnosis.
    \item Optimization of training parameters using depthwise separable convolution. The training parameters of the MAG-Net reduced from 26.0M to 5.4M.
    \item MAG-Net achieved significant improvement in classification and segmentation as compared to the state-of-the-art models while having limited data samples.
\end{itemize}
The rest paper is organized as follows: Section 2 describes the crux of related work on brain tumor segmentation and classification. Section 3, talks about the proposed architecture, whereas Section 4 discuses the training and testing environment with experimental and comparative analysis. Finally, concluding remarks are presented in Section 5.

\section{Literature review}
Identifying the brain tumor is a challenging task for the radiologists. Recently, several deep learning based approaches are proposed to aid in faster diagnosis of the diseases. Segmentation of the infected region is most common and critical practice involved in the diagnosis. In addition, the segmented region can be provided with label (classification) to indicate what type of anomaly or infection is present in the image. 

In contrast to the traditional approaches, Cheng et al.~\cite{cheng2015enhanced} proposed a brain tumor classification approach using augmented tumor region instead of original tumor region as RoI (region of interest). Authors utilized the bag of word (BOW) technique to segment and extract local features from RoI. Dictionary is used for encoding the extracted local features maps that are passed through SVM (support vector machine) classifier. The approach outperformed the traditional classification techniques with the accuracy of 91.28\% but the performance is limited by the data availability. In similar work, Ismael et al.~\cite{ismael2018brain} proposed an approach of combining statistical features along with neural networks by using filter combination: discrete wavelet transform (DWT)(represented by wavelet coefficient) and Gabor filter (for texture representation). For classification of the tumor, three layered neural network classifier is developed using multilayer perceptron network that is trained with statistical features. In contrast to Cheng et al.~\cite{cheng2015enhanced}, authors also achieved promising results on the limited data samples with an overall accuracy of. 91.9\%. 

Recently, capsule network~\cite{hinton2018matrix} has shown great performance in many fields especially in biomedical image processing. Afshar et al.~\cite{afshar2018brain} proposed basic capsnet with three capsules in last layer representing three tumor classes. However, due to varied behavior (background, intensity, structure, etc.) of MRI image, the proposed model failed to extract optimal features representing the tumor structure. The author achieved the tumor classification accuracy of 78\% and 86.5\% using raw MRI images and tumor segmented MRI images respectively. In another approach, Pashaei et al.~\cite{pashaei2018brain} utilized CNN and kernel extreme learning machine that comprises one hidden layer with 100 neurons to increase the robustness of the model. With several experimental trials, the authors achieved an accuracy of 93.68\% but detects only 1\% of the positive pituitary tumor cases out of the total pituitary tumor case. Deepak et al.~\cite{deepak2019brain} proposed a transfer learning approach that uses pre-trained GoogleNet model to extract features (referred as deep CNN features) with softmax classifier in the output layer to classify three tumor classes. Furthermore, the authors combine the deep CNN features and SVM model to analyse the classification performance. The authors achieved 97.1\% accuracy but resulted in poor performance by standalone GoogleNet model due to overfitting with limited training image dataset, and misclassifications in meningioma tumor. In another approach, Pernas et al.~\cite{pub.1135094000} proposed to process images in three different spatial scales along with multi pathways feature scales for classification and segmentation of brain tumor. The images are pre-processed with elastic transform for preventing overfitting. The model analyses entire image and classifies pixel by pixel in one of four possible output labels (i.e. 0-healthy, 1-meningioma, 2-glioma, and 3-pituitary tumor). The proposed approach outperformed existing approaches with 97.3\% classification accuracy, but with poor segmentation performance. Following this context, in this article multi-task attention guided network (MAG-Net) is proposed based on the U-Net architectural design ~\cite{ronneberger2015u} that uses parallel depthwise separable convolution layers for multi-level feature extraction along with an attention mechanism to better extract tumor features for brain tumor classification and generate the corresponding tumor mask.


 \begin{figure}[]
    \centering
    \includegraphics[scale=0.6]{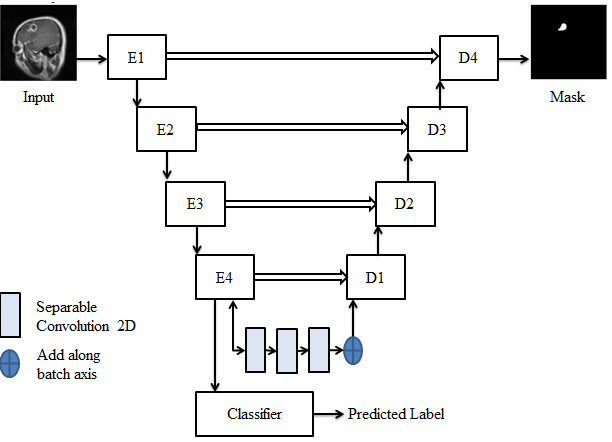}
    \caption{Schematic representation of the architecture of MAG-Net model.}
    \label{fig:model}
\end{figure}

\section{Proposed work}
The proposed multi-task attention guided network (MAG-Net) model, as shown in Fig.~\ref{fig:model}, focuses on reducing overall computation, better feature extraction and optimizing the training parameters by reduction. The overall architectural design consists of an encoder, decoder, and classification module with 5.4M trainable parameters. The overall architectural design of the model is inspired by the U-Net encoder-decoder style~\cite{punn2020multi}. Due to its state-of-the-art potential, this model is the most prominent choice among the researchers to perform biomedical image segmentation~\cite{minaee2021image}.

In MAG-Net to reduce the number of training parameters without the cost of performance, standard convolution operations are replaced with depthwise separable convolution. In addition, the skip connections are equipped with attention filters~\cite{oktay2018attention} to better extract the feature maps concerning the tumor regions. The attention approach filters the irrelevant feature maps in the skip connection by assigning weights to highlight its importance towards the tumor regions. Besides, the encoder block is equipped with parallel separable convolution filters of different sizes, where the extracted feature maps are concatenated for better feature learning. These features are then passed to the corresponding decoder blocks via attention enabled skip connections to aid in feature reconstruction with the help of upsampling operation. The bottleneck layer connects the feature extraction path to the feature reconstruction path. In this layer filters of different sizes are used along with the layer normalization. Furthermore, the classification is performed using the extracted feature maps obtained from the final encoder block.


\subsection{Encoder} 
\begin{figure}[]
    \centering
    \includegraphics[scale=0.45]{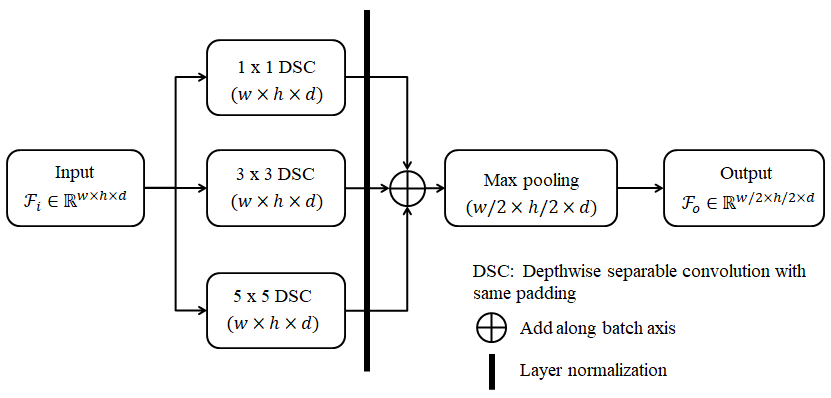}
    \caption{Proposed MAG-Net 2D encoder module.}
    \label{fig:encoder}
\end{figure}

To detect the shape and size of varying image like brain tumor it is required to use separable convolution of different sizes. Inspired from the concept of inception neural network~\cite{punn2020inception} the encoder segment is consist of separable convolutions of 1 x 1, 3 x 3, and 5 x 5 kernels. Each of separable convolutions are followed by layer normalization. The extracted feature maps are fused with add operation that are downsampled by max pooling operation. Fig.~\ref{fig:encoder}, shows the proposed encoder architecture of MAG-Net model for some input feature map, $\mathcal{F}_i \in \mathcal{R}^{w\times h\times d}$, where $w$, $h$ and $d$ are the width, height and depth of the feature map.

\subsection{Decoder} The decoder component follows from the encoder block and that tend to reconstruct the spatial dimension to generate the output mask in same dimension as input. It consists of upsampling of the feature maps along with the concatenation with the attention maps followed by a separable convolution operation. Long skip connections~\cite{drozdzal2016importance} are used to propagate the attention feature maps from encoder to decoder to recover spatial information that was lost during downsampling in encoder. By using attention in the skip connection it helps the model to suppress the irrelevant features.

\subsection{Classification} This module classifies the brain tumor MRI images into respective classes i.e meningioma, glioma, and pituitary tumor by utilizing the features extracted from the encoder block. This encoder block act as backbone model for both classification and segmentation, thereby reducing the overall complexity of the model. In this classification block the feature maps of the last encoder block act as input that are later transformed into 1D tensor by using global average pooling. The pooled feature maps are then processed with multiple fully connected layers. The classification output is generated from the softmax activated layer that generates the probability distribution of the tumor classes for an image. 

\section{Experiment and Results}

\subsection{Dataset Setup}
\begin{figure}[]
    \centering
    \includegraphics[scale=0.6]{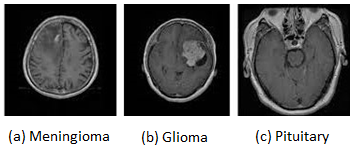}
    \caption{A slice of MRI scan with T1 modality showing different tumor classes: meningioma, glioma, and pituitary}
    \label{fig:DS}
\end{figure}

The present research work utilizes Figshare~\cite{cheng2017} dataset that comprises of 2D MRI scan with T1-weighted contrast-enhanced modality acquired from 233 patients to form a total of 3064 MRI scans. The T1 modality highlight distinct features of the brain tumor with three classes representing the type of brain tumor i.e. meningioma (708 slices), glioma (1426 slices), and pituitary (930 slices) forming 23\%, 46.5\%, and 30\% class distribution in the dataset respectively. The sample MRI slices of different tumor classes are presented in Fig.~\ref{fig:DS}. Dataset is randomly split into 80\% training and 20\% of the validation set. The training and testing composition kept the same throughout the experiment trails for comparative analysis.

\subsection{Training and Testing}
The MAG-Net model is trained and evaluated on the Figshare dataset. The training phase is accompanied with early-stopping~\cite{Brownlee:} to tackle the overfitting problem, and Adam as a learning rate optimiser ~\cite{ruder2017overview}. Cross entropy based loss functions are most popularly used for model training and validating segmentation and classification tasks. Following this, binary cross entropy and categorical cross entropy functions are employed for training the model for binary tumor mask generation and classification respectively. Binary cross entropy (BCE, shown in Eq.~\ref{eqn1}) is a sigmoid activation ~\cite{jamel2012implementation} followed by cross entropy loss~\cite{zhang2018generalized} that compares each of the predicted probabilities to actual output. Categorical cross entropy (CE, shown in Eq.~\ref{eqn2}) is a softmax activation function followed by cross-entropy-loss that compares the output probability over each tumor class for each MRI image.  

\begin{equation} \label{eqn1}
\mathcal{L}_{BCE}  = -\frac{1}{N} \sum_{i = 1}^N(y_i. log(p(y_i)) + (1-y_i).log(1 - P(y_i)))
\end{equation}
where $y$ represents actual tumor mask, $p(y)$ represents predicted tumor mask and $N$ is the total number of images. 

\begin{equation} \label{eqn2}
\mathcal{L}_{CE} = \sum_i^C t_i log(f(s_i)) 
\end{equation}
where $C$ is the no. of class, $f(s_i)$ is the probability of occurrence of each class $t_i$ represents 1 for true label and 0 for others.

For segmentation the most popular evaluation matrics are dice coefficient (shown in Eq.~\ref{eqn3}) and intersection-over-union (IoU / Jaccard index) (shown in Eq.~\ref{eqn4}), and hence are utilized to evaluate the trained MAG-Net model. TP defines correctly classified predictions FP defines wrongly classified, and FN defines missed objects of each voxel. 

\begin{equation} \label{eqn3}
Dice Coefficient = \frac{2 * TP}{2 * TP + FP + FN}
\end{equation}

\begin{equation} \label{eqn4}
IoU = \frac{ TP}{TP + FP + FN}
\end{equation}

To evaluate classification module of the MAG-Net, accuracy, precision, recall, f1-score and micro average metrics are considered for better quantification and visualization of the performance of the model. Precision of the class, as shown in Eq.~\ref{eqn5}, quantifies about the positive prediction accuracy of the model. Recall is the fraction of true positive which are classified correctly (shown in Eq.~\ref{eqn6}). F1-score quantifies the amount of correct predictions out of all the positive predictions (shown in Eq.~\ref{eqn7}). Support quantifies the true occurrence in the specified dataset of the respective class. Micro average ($\mu _{avg}$) (shown in Eq.~\ref{eqn8}, Eq.~\ref{eqn9} and Eq.~\ref{eqn10}) is calculated for precision, recall, and F1-score. To compute micro average ($\mu _{avg}$), the test dataset is divided into two sub dataset, on each of which the true positive, false positive and false negative predictions are identified.   

\begin{equation}\label{eqn5}
Precision = \frac{TP}{(TP + FP)}
\end{equation}

\begin{equation}\label{eqn6}
Recall = \frac{TP}{(FN + FP)}
\end{equation}

\begin{equation}\label{eqn7}
F1-score = \frac{2 * Recall * Precision}{(Recall + Precision)}
\end{equation}

\begin{equation}\label{eqn8}
\mu _{avg}(Precision) = \frac{TP_1 + TP_2}{(TP_1 + TP_2 + FP_1 + FP_2)}
\end{equation}

\begin{equation}\label{eqn9}
\mu _{avg}(Recall) = \frac{TP_1 + TP_2}{(TP_1 + TP_2 + FN_1 + FN_2)}
\end{equation}

\begin{equation}\label{eqn10}
\mu _{avg}(F1-score) = HM(\mu _{avg}(Precision), mu _{avg}(Recall))
\end{equation}
where $TP_1$, $FP_1$, and $FN_1$ belong to the first set and $TP_2$, $FP_2$, and $FN_2$ belongs to the different sets. $HM$ is the harmonic mean. 

\begin{figure}[]
    \centering
    \includegraphics[scale=0.7]{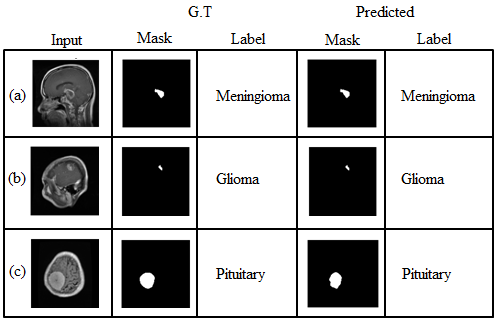}
    \caption{Qualitative results of brain tumor segmentation and classification on MRI images (a, b, and c of different tumor classes) using MAG-Net model.}
    \label{fig:fig5}
\end{figure}

\subsection{Results}
The MAG-Net outputs the segmented mask of a given MRI image consisting of tumor region corresponding to meningioma, glioma, and pituitary as classified by the model. For randomly chosen MRI slices,  Fig.~\ref{fig:fig5} presents the segmentation and classification results of model. The visual representation confirms that the results are close to the ground truth of respective tumor classes.

\begin{table}[]
    \centering
    \caption{\label{tab:table-2} Comparative analysis of the MAG-Net with the existing segmentation models on test dataset.} 

 \begin{tabular}{|l||l|l|l|l|l|}
 \hline
 \textbf{Model}& \textbf{Accuracy}& \textbf{Loss} &\textbf{Dice coefficient} & \textbf{Jaccard index} & \textbf{Parameters} \\
 \hline
 \hline
 U-Net  &99.5 & 0.024 & 0.70 & 0.55 & 31M \\\hline
 wU-Net &99.4 & 0.034 & 0.66 & 0.49 & 31M\\ \hline
 Unet++ &99.5 & 0.028 & 0.65 & 0.49 & 35M\\\hline
 MAG-Net   &\textbf{99.52}&\textbf{0.021} &  \textbf{0.74} & \textbf{0.60} &\textbf{5.4M}\\
 \hline
 \multicolumn{6}{l}{*bold quantities indicate the best results.}
\end{tabular}    
\end{table}

Table \ref{tab:table-2} represents the result of the proposed work for segmentation in the form of accuracy, loss, dice coefficient, Jaccard index, and trainable parameters along with comparative analysis with other popular approaches. The proposed framework outperforms the other approaches in segmenting tumor with the dice and IoU score of 0.74 and 0.60 respectively. In contrast to other models, MAG-Net achieved best results with minimal trainable parameters. The other popular approaches taken in comparative analysis for segmentation are U-Net~\cite{Adityajain_git:}, U-Net++~\cite{U-Net++,zhou1807nested}, and wU-Net.~\cite{Nested-U-Net}. 

\begin{table}[h!]
    \centering
    \caption{\label{tab:table-3} Comparative analysis of the MAG-Net with the existing classification models on test dataset using confusion matrix.} 
\begin{tabular}{|p{1.75cm}|p{1.25cm}|p{1.25cm}||p{2cm}|p{2.1cm}|p{1.5cm}|p{1.5cm} | }
 \hline
 \textbf{Model}& \textbf{Acc.}& \textbf{Loss} & &\textbf{Meningioma} & \textbf{Glioma} & \textbf{Pituitary} \\
 \hline
 \hline
    & & & Meningioma &114 &25 &3 \\ \cline{4-7}
 VGG16 &93.15& 0.26 & Glioma &13 &271 &1\\ \cline{4-7}
     & & & Pituitary &0 &1 &185 \\ \hline
   & & & Meningioma &114 &21 &7 \\ \cline{4-7}
 VGG19 & 93.8 & 0.25& Glioma &11 &274 &0\\ \cline{4-7}
     & & & Pituitary &0 &1 & 185\\ \hline
   & & & Meningioma &123 &12 &7 \\ \cline{4-7}
 ResNet50 & 94.2 & 0.31& Glioma &16 &266 &3\\ \cline{4-7}
     & & & Pituitary & 1 & 0&185 \\ \hline
   & & &  Meningioma &\textbf{134} & \textbf{7}&\textbf{1} \\ \cline{4-7}
 MAG-Net&\textbf{98.04}& \textbf{0.11} & Glioma &\textbf{1} &\textbf{282} &\textbf{2}\\ \cline{4-7}
   &    & & Pituitary &\textbf{1} &\textbf{0} &\textbf{185} \\ \cline{4-7}
 \hline
 \multicolumn{7}{l}{*bold quantities indicate the best results.}
\end{tabular}
\end{table}

Table~\ref{tab:table-3} and Table~\ref{tab:table-4} represent the results of the proposed work for classification in the form of accuracy, loss, confusion matrix, and classification report for meningioma, glioma, and pituitary tumor along with comparative analysis with other state-of-the-art approaches: VGG-16~\cite{VGG16}, VGG-19~\cite{VGG16}, and ResNet50~\cite{resnet}. With exhaustive experimental trials it is observed that MAG-Net outperformed the existing approaches with significant margin in all the metrics.

\begin{table}
    \centering
        \caption{\label{tab:table-4} Comparative analysis of the MAG-Net with the existing classification models on test dataset considering classification report as evaluation parameter.} 
\begin{tabular}{|p{2cm}||p{2.15cm}|p{2cm}|p{1.85cm}|p{2cm}|p{1.5cm} | }
 \hline
 \textbf{Model}& \textbf{Classes} & \textbf{Precision} & \textbf{Recall} & \textbf{f1-Score} & \textbf{Support} \\
 \hline
 \hline
  \multirow{4}{*}{VGG16}  &  Meningioma& 0.90 &0.80 & 0.85 &142 \\ \cline{2-6}
  & Glioma & 0.91 & 0.85 & 0.93 &285\\ \cline{2-6}
      & Pituitary & 0.98&0.99 & 0.99 &186 \\ \cline{2-6}
     &Micro avg. & 0.93 & 0.93 &  0.93 &613\\ \hline
 \multirow{4}{*}{VGG19}&  Meningioma& 0.91 &0.80 & 0.85 &142 \\ \cline{2-6}
  & Glioma & 0.93 & 0.96 & 0.94 &285\\ \cline{2-6}
      & Pituitary & 0.96&0.99 & 0.98 &186 \\ \cline{2-6}
     &Micro avg. & 0.93 & 0.93 &  0.93 &613\\ \hline
 \multirow{4}{*}{ResNet-50}&  Meningioma& 0.88 &0.87 & 0.87 &142 \\ \cline{2-6}
  & Glioma & 0.93 & 0.99 & 0.94 &285\\ \cline{2-6}
      & Pituitary & 0.95&0.99 & 0.97 &186 \\ \cline{2-6}
     &Micro avg. & 0.94 & 0.94 &  0.94 &613\\ \hline
 \multirow{4}{*}{MAG-Net}&  Meningioma& \textbf{0.99} &\textbf{0.94} & \textbf{0.96} &142\\ \cline{2-6}
  & Glioma & \textbf{0.98} & \textbf{0.99} & \textbf{0.98} &285\\ \cline{2-6}
     & Pituitary & \textbf{0.98} &\textbf{0.99} & \textbf{0.99} &186 \\ \cline{2-6}
     &Micro avg.& \textbf{0.98} & \textbf{0.98} &  \textbf{0.98} &613\\ \hline

 \multicolumn{6}{l}{*bold quantities indicate the best results.}
\end{tabular}    
\end{table}

It is observed that unlike other state-of-the-art models, the MAG-Net model achieved promising results due to the reduction in the overall computation, better feature extraction and training parameters optimization. As shown in Table~\ref{tab:table-2} raw U-Net displayed similar performance but at the cost  of large number of trainable parameters. In the MAG-Net model, the encoder block is developed by replacing convolution layers with parallel depthwise separable convolution of various sizes connected in parallel which resulted in better multi-scale feature learning for varying shapes and sizes of the tumor. For reducing spatial loss during feature reconstruction, attention mechanism is used in skip connections for better feature reconstruction. To reduce the overall complexity of the model the feature extracted by encoder blocks are reused to classify the type of brain tumor.

\section{Conclusion}
In this paper, the complex task of brain tumor segmentation and classification is addressed using multi-task attention guided network (MAG-Net). This a U-Net based model that features reduction in the overall computation, better feature extraction and training parameters optimization. The proposed architecture achieved significant performance on the Figshare brain tumor dataset by exploiting the state-of-the-art advantages of U-Net, depthwise separable convolution and attention mechanism. The MAG-Net model recorded the best classification and segmentation results compared to the existing classification and segmentation approaches. It is believed that this work can also be extended to other domains involving classification and segmentation tasks.

\section*{Acknowledgment}
	We thank our institute, Indian Institute of Information Technology Allahabad (IIITA), India and Big Data Analytics (BDA) lab for allocating the centralised computing facility and other necessary resources to perform this research. We extend our thanks to our colleagues for their valuable guidance and suggestions.

\bibliographystyle{splncs04}
\bibliography{reference}

\end{document}